\documentclass[preprint,showpacs,preprintnumbers,amsmath,amssymb,aps]{revtex4-1}
\usepackage{graphicx}
\usepackage{dcolumn}
\usepackage{bm}

\begin{document}

\title{Hole doping, hybridization gap, and electronic correlation in graphene on a platinum substrate}

\author{Jinwoong Hwang}
\affiliation{Department of Physics, Pusan National University, Busan 46241, Korea}
\author{Hwihyeon Hwang}
\affiliation{Department of Physics, Pusan National University, Busan 46241, Korea}
\author{Min-Jeong Kim}
\affiliation{Department of Physics, Pusan National University, Busan 46241, Korea}
\author{Hyejin Ryu}
\affiliation{Advanced Light Source, Lawrence Berkeley National Laboratory, Berkeley, CA 94720, USA}
\author{Ji-Eun Lee}
\affiliation{Department of Physics, Pusan National University, Busan 46241, Korea}
\author{Qin Zhou}\email{zhou@unl.edu}
\affiliation{Mechanical and Materials Engineering, University of Nebraska, Lincoln, Nebraska 68588, USA}
\author{Sung-Kwan Mo}
\affiliation{Advanced Light Source, Lawrence Berkeley National Laboratory, Berkeley, CA 94720, USA}
\author{Jaekwang Lee}
\affiliation{Department of Physics, Pusan National University, Busan 46241, Korea}
\author{Alessandra Lanzara}
\affiliation{Department of Physics, University of California, Berkeley, CA 94720, USA}
\author{Choongyu Hwang}\email{ckhwang@pusan.ac.kr}
\affiliation{Department of Physics, Pusan National University, Busan 46241, Korea}

\date{\today}

\begin{abstract}
The interaction between graphene and substrates provides a viable routes to enhance functionality of both materials. Depending on the nature of electronic interaction at the interface, the electron band structure of graphene is strongly influenced,  allowing us to make use the intrinsic properties of graphene or to design additional functionality in graphene. Here, we present an angle-resolved photoemission study on the interaction between graphene and a platinum substrate. The formation of an interface between graphene and platinum leads to a strong deviation in the electronic structure of graphene not only from its freestanding form but also from the behavior observed on typical metals. The combined study on the experimental and theoretical electron band structure unveils the unique electronic properties of graphene on a platinum substrate, which singles out graphene/platinum as a model system investigating graphene on a metallic substrate with strong interaction.
\end{abstract}

%
%
%
%
%

\maketitle

\section{Introduction}

The interface between graphene and transition-metals has been widely investigated not only to prepare transferrable high quality graphene, but also to provide a versatile platform for device applications. For example, transition-metals such as Cu provide an excellent opportunity to grow wafer-size graphene that can be transferred onto insulating substrates~\cite{Hong}, which has potential applications such as flexible displays~\cite{Hong2} and transparent electrodes~\cite{Nobis}. In addition, ferromagnetic substrates such as Co result in a mini Dirac cone that consists of single spin, which originates from the hybridization between graphene $\pi$ and Co 5$d$ bands~\cite{Usachov}. Alternatively, graphene itself can play an important role in enhancing functionality of other materials or devices such as the electro-catalytic effect of Pt~\cite{Serger,Sun} or solar cells as a counter electrode~\cite{Lee}. The key factor determining the nature of such cooperation is electronic interactions between graphene and the transition-metals~\cite{TransitionMetals}.

However, despite intense studies on graphene/transition-metals, the effect of some of the transition-metals on the electronic properties of graphene is still unclear. For example, a Raman spectroscopy study on the interface between graphene and Pt reports strongly suppressed Raman signals from graphene signifying strong interactions between them~\cite{Qin}, which can possibly cause Rashba-type spin splitting in graphene~\cite{Ilya1,Ilya2}. On the other hand, a recent experimental study on the electron band structure of graphene on a Pt(111) surface shows that overlying graphene exhibits typical characteristics of free-standing graphene~\cite{Shuyun,Sutter2}. However, first-principles calculations support strong interactions between graphene and Pt~\cite{Giovannetti,Wang}. 

The controversy can be settled down when the electron band structure of both graphene and the Pt substrate are measured simultaneously using angle-resolved photoemission spectroscopy (ARPES). ARPES is a powerful technique to understand not only the fundamental electronic properties of a solid state material via the direct measurement of its electron band structure, but also charge carrier dynamics through the analysis of electron self-energy~\cite{Damascelli}. As a result, ARPES is expected to unveil information on the nature of the electronic interaction between graphene and the Pt substrate.

In this report, we study the electron band structure of graphene on a polycrystalline Pt foil. We observe strong interaction between graphene and the Pt substrate resulting in hybridization gaps and hole doping in the graphene $\pi$ band. These findings are very different from the previous ARPES results on graphene/Pt(111)~\cite{Shuyun,Sutter2} that report freestanding nature of  graphene, but consistent with first-principles calculation for the same system~\cite{Giovannetti,Wang}. Concomitantly, the $\pi$ band of graphene on Pt differs from that of graphene on other metallic substrates as discussed below~\cite{SiegelCu,Varykhalov,Rader,Sutter}. 

\section{Experimentals}

Graphene samples were prepared on a 0.1~mm thick polycrystalline Pt foil using the chemical vapor deposition (CVD) method~\cite{Qin}. The substrate was annealed at 950~$^{\circ}{\rm C}$ during the growth and transferred to a ultra-high vacuum chamber followed by cleaning process via e-beam heating upto 850~$^{\circ}{\rm C}$ to remove air contaminants. For typical metals whose crystal has the {\it fcc} structure such as Pt, the (111) orientation is energetically preferable after thermal treatment~\cite{Bodepudi,Kang,Shintani} (see Supplementary Information). During the cleaning, the chamber pressure was under $4\times10^{-9}$~Torr. ARPES experiments have been performed at the beamline 10.0.1 of the Advanced Light Source at Lawrence Berkeley National Laboratory. All the data have been measured with a photon energy of 50~eV and the sample temperature during the measurements was 20~K. Energy and momentum resolutions throughout the experiments were 24~meV and 0.04~\AA$^{-1}$, respectively. 

The electron band structure calculations have been carried out using density-functional theory with the plane-wave based Vienna {\it ab initio} package (VASP)~\cite{Kresse}. The projector-augmented wave method was used to mimic the ionic cores, while the generalized gradient approximation (GGA) was employed for the exchange-correlation functional~\cite{Perdew}. We used a kinetic energy cut-off of 500~eV and $\Gamma$-centered $8\times8\times2$ k-point meshes for the Brillouin zone integration. The calculations are converted in energy to $10^{-6}$~eV/cell, and the structures are relaxed until the forces are less than $5\times10^{-3}$~eV/\AA. For graphene, we find a lattice constant of 2.46~\AA. Five layers of Pt atoms are considered as the Pt(111) surface. The hexagonal graphen/Pt supercell is constructed with a $2\times2$ graphene unit cell adsorbed on one side of the $\sqrt{3}\times\sqrt{3}$ unit cell of the Pt(111) surface, and its in-plane lattice constant is fixed by a $2\times2$ graphene lattice parameter. The distance between graphene and Pt(111) surface is chosen as 3.3~\AA~ because the theoretical band structure calculated using 3.3~\AA~ reproduces experimentally measured band structure very well. It is also reported that the separation of 3.3~\AA~reproduces the x-ray reflectivity data~\cite{Hass}. In order to avoid spurious interaction between images of the supercell in the [001] direction, a vacuum of 28~\AA~is considered.

\section{Results and Discussions}

\begin{figure*}[t]
\centering
\includegraphics[width=0.9\columnwidth]{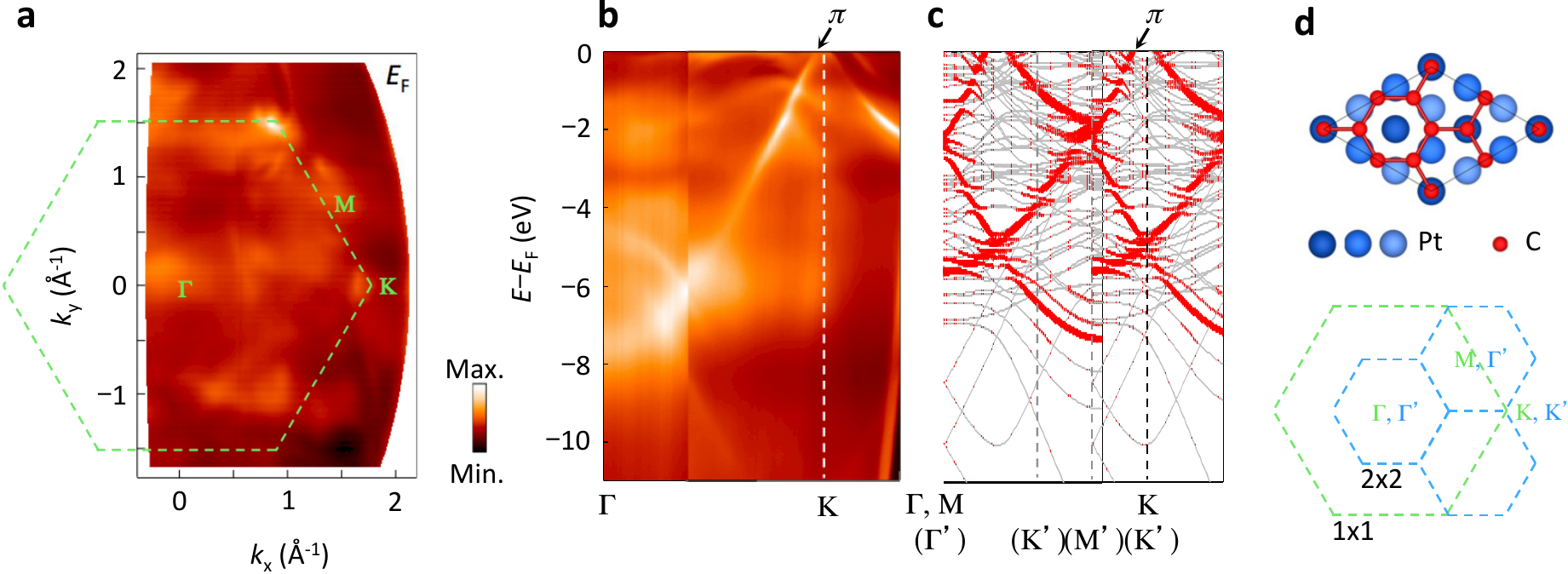}
\caption{\label{fig:fig1} {\bf a}, Fermi surface of graphene/Pt. Green dashed-hexagon denotes the graphene unit cell. {\bf b}, An ARPES intensity map taken along the $\Gamma{\rm K}$ direction of the graphene unit cell denoted in panel {\bf a}. {\bf c}, Calculated electron band structure of graphene/Pt. $\Gamma^{\prime}$, K$^{\prime}$, and M$^{\prime}$ denote the high symmetry points of the graphene/Pt supercell, when $\Gamma$, K, and M denote those of the graphene unit cell. {\bf d}, Top-view of a structural model of graphene/Pt. Lighter blue spheres denote Pt atoms in the upper layer. The lower panel shows the graphene $1\times1$ unit cell (green dashed-hexagon) and $2\times2$ supercell (blue dashed-hexagon).}
\end{figure*}

Figure~1{\bf a} shows a constant energy intensity map of graphene/Pt taken at $E_{\rm F}$, when the green dashed-hexagon denotes the first Brillouin zone of the graphene unit cell. Instead of the typical spot-like intensity distribution at the Brillouin zone corner, K point, of graphene~\cite{Aaron}, the measured Fermi surface consists of complicated intensity pattern. To find the signature of graphene, an ARPES intensity map was taken along the $\Gamma$K direction of the graphene unit cell as shown in Fig.~1{\bf b}. One can find a dispersive band near the $\Gamma$ point with a band minimum at $\sim$8~eV below Fermi energy, $E_{\rm F}$, that approaches toward $E_{\rm F}$ near the K point. This resembles the common feature of graphene $\pi$ band. However, deviations from the typical $\pi$ band dispersion are clear in that the measured band does not show the top of the conical dispersion around the K point but exhibits unusual intensity variation around 1$\sim$2~eV below $E_{\rm F}$. To characterize the measured ARPES intensity, the electron band structure of graphene/Pt was calculated as shown in Fig.~1{\bf c}. Due to the presence of the Pt substrate, the supercell including both graphene and the Pt substrate was taken into account as shown in the upper panel of Fig.~1{\bf d}, resulting in the graphene $\pi$ band in the 1$\times$1 unit cell to be folded into the 2$\times$2 supercell as shown in the lower panel of Fig.~1{\bf d}. As a result, the $\Gamma$ and M points of the graphene 1$\times$1 unit cell lie on the Brillouin zone center of the 2$\times$2 supercell, $\Gamma^{\prime}$, while the K point lies on the Brillouin zone corner K$^{\prime}$. In the calculated band structure, the red curves denote the electronic states with a graphene 2$p_{\rm z}$ orbital character and the grey curves all the other electronic states of both graphene and Pt. Considering the folded scheme of the band structure, the comparison of the measured (Fig.~1{\bf b}) and the calculated (Fig.~1{\bf c}) bands indicates that the measured dispersive band originates from the graphene $\pi$ band as denoted by a black arrow in each panel.

\begin{figure*}
\centering
\includegraphics[width=0.5\columnwidth]{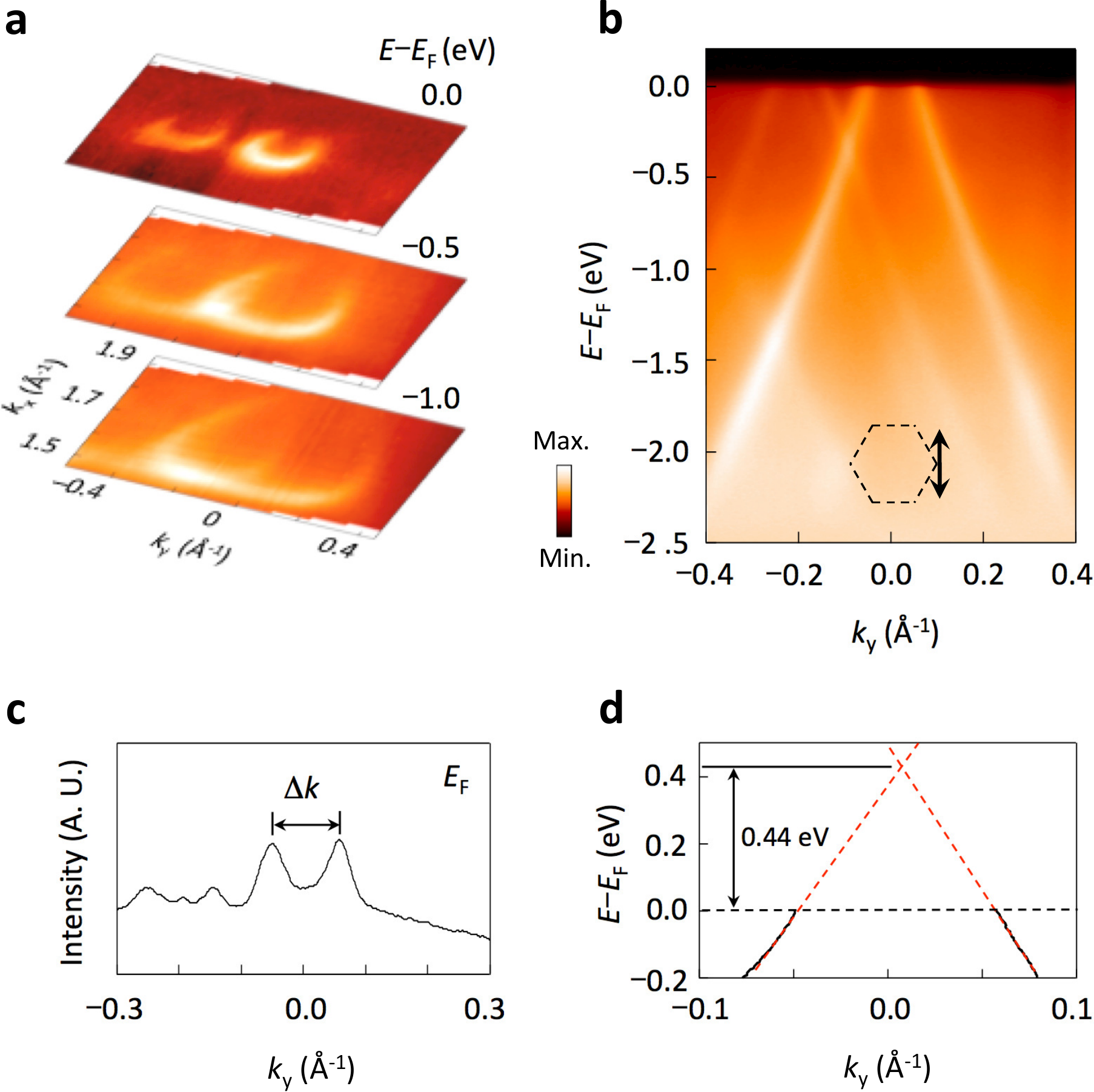}
\caption{\label{fig:fig2} {\bf a}, Constant energy maps around the K point taken at $E-E_{\rm F} = $0.0, $-$0.5, and $-$1.0~eV. Two conical dispersions indicate that ARPES probes two pieces of graphene with slight azimuthal misorientation. {\bf b}, An ARPES intensity map near $E_{\rm F}$ taken across the K point perpendicular to the $\Gamma{\rm K}$ direction of the graphene unit cell as denoted in the inset. {\bf c}, A momentum distribution curve at $E_{\rm F}$. $\Delta\,k$ denotes the separation of the two branches of the graphene $\pi$ bands with relatively strong spectral intensity at $E_{\rm F}$. {\bf d}, The energy-momentum dispersion of the bands near $E_{\rm F}$ obtained using a Lorentzian fit function. Extended straight lines give a naive estimation of the Dirac energy, e.\,g.\,, 0.44~eV above $E_{\rm F}$.}
\end{figure*}

The detailed ARPES intensity map near $E_{\rm F}$ gives fundamental information on the $\pi$ band of graphene on a Pt substrate. Figure~2{\bf a} shows constant energy intensity maps around the K point taken at three different energies relative to $E_{\rm F}$ that are $0$, $-0.5$, and $-1.0$~eV. At $E-E_{\rm F}=0~{\rm eV}$, the constant energy map shows two crescent-like intensity distributions, each of which originates from the pseudospin nature of graphene~\cite{HwangPRB}. The two crescent-like shapes indicate the presence of two relatively wide graphene sheets out of multiple pieces with azimuthal disorder, typical of graphene prepared using the CVD method~\cite{Walter}, within a photon beam spot of $40\times80~\mu{\rm m}^2$ in the ARPES measurements.  With decreasing $E-E_{\rm F}$, the crescent-like shape gradually increases, indicating that there exists a conical dispersion of graphene. Figure~2{\bf b} shows an ARPES intensity map taken along the $k_{\rm y}$ direction at $k_{\rm x}$=1.7~{\rm \AA}$^{-1}$, i.\,e.\,, perpendicular to the $\Gamma$K direction as denoted in the inset. Strong and weak spectral intensities are observed due to the presence of two pieces of graphene with slightly different azimuthal orientations as discussed in Fig.~2{\bf a}. 

It is important to note that the measured ARPES intensity map does not show the Dirac energy, $E_{\rm D}$, where the conduction and valence bands of graphene meet at a single point. This indicates that $E_{\rm D}$ exists above $E_{\rm F}$, providing a direct evidence of charge transfer from graphene to the Pt substrate compared to freestanding graphene where $E_{\rm D}$ exactly aligns to $E_{\rm F}$. The amount of charge transfer is determined by the distance between the two branches of the conical dispersion in the momentum distribution curve (MDC) taken at $E_{\rm F}$, i.\,e.\,, $\Delta\,k$ shown in Fig.~2{\bf c}, resulting in a hole carrier density of $0.91\times10^{13}~{\rm cm}^{-2}$. Alternatively, the amount of hole doping can be estimated by the position of $E_{\rm D}$ relative to $E_{\rm F}$ as shown in Fig.~2{\bf d}. Extended straight lines over the graphene $\pi$ band taken by a Lorentzian fit to the MDCs give $E_{\rm D}$ of 0.44~eV above $E_{\rm F}$.

\begin{figure}[t]
\centering
\includegraphics[width=0.5\columnwidth]{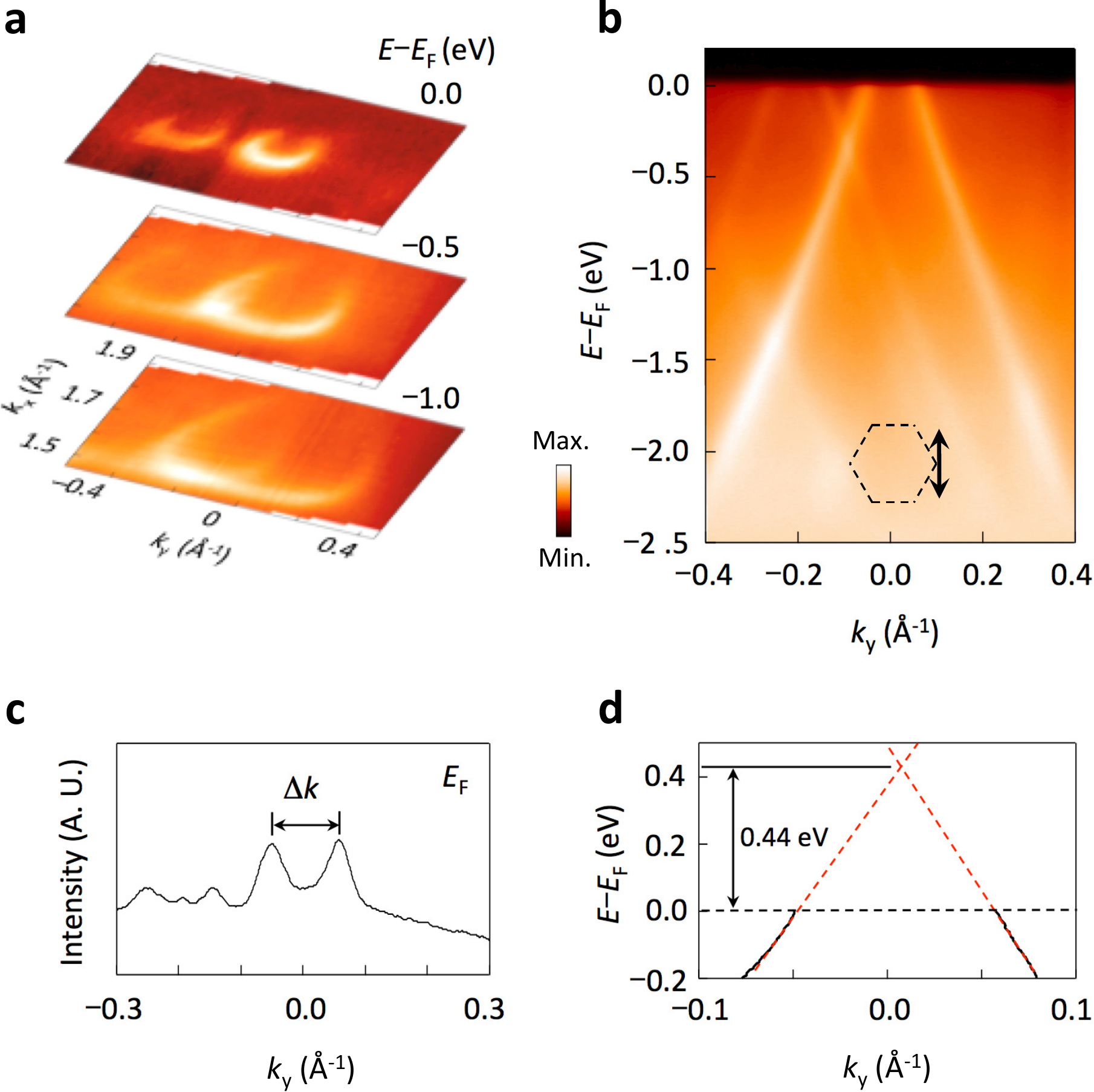}
\caption{\label{fig:fig3} {\bf a}-{\bf b}, An ARPES intensity map and its first derivative near $E_{\rm F}$ taken across the K point along the $\Gamma$K direction as denoted in the inset. {\bf c}-{\bf d}, Calculated electron band structure of graphene/Pt. Red and blue dots denote C 2$p_{\rm z}$ and Pt 5$d$ orbitals, respectively.}
\end{figure}

A comparison of the measured and calculated bands of graphene/Pt near $E_{\rm F}$ taken along the $\Gamma$K direction provides additional information on the electronic interaction between graphene and the Pt substrate. As shown in Fig.~3{\bf a}, the $\pi$ band exhibits unusual intensity variation compared with freestanding graphene or slightly hole-doped graphene~\cite{Pletikosic,HwangSR,Sprinkle,Siegel,Johansson}. The first derivative of the ARPES intensity map shown in Fig.~3{\bf b} exhibits clear discontinuities in both intensity and energy-momentum dispersion at the crossing points with other bands as denoted by green dashed-ovals and arrows. Such intensity variation and discontinuities in the dispersion indicate that there exists a strong hybridization among different bands~\cite{HwangYb,HwangJKPS}.

To find out the origin of the additional bands that deform the graphene $\pi$ band, the measured data are compared to the calculated band structure taken along the $\Gamma^{\prime}$K$^{\prime}$ direction of the 2$\times$2 supercell, i.\,e.\,, the $\Gamma$K direction of the graphene unit cell, as shown in Figs.~3{\bf c} and~3{\bf d}. Here red and blue dots denote the electron band structure with graphene 2$p_{\rm z}$ and Pt 5$d$ orbital characters, respectively, and bigger dots correspond to stronger spectral intensity. The graphene $\pi$ band exhibits clear discontinuities at the crossing points with the Pt 5$d$ bands. Especially, two Pt 5$d$ bands indicated by green arrows reproduce the additional bands observed in Figs.~3{\bf a} and~3{\bf b}, although single crystal Pt is used for the band structure calculation instead of polycrystalline Pt. In the calculated band structure, the Pt 5$d$ bands also exhibits clear discontinuities at the crossing points with the graphene $\pi$ bands. In addition, each graphene 2$p_{\rm z}$ and Pt 5$d$ orbital plotted in Figs.~3{\bf c} and~3{\bf d} shows weak spectral intensity following the trace of each other, respectively. The comparison of the measured and calculated bands gives a clear signature of hybridization between graphene $\pi$ and Pt 5$d$ bands, and resultant hybridization gaps in each material.

\begin{figure}[t]
\centering
\includegraphics[width=0.5\columnwidth]{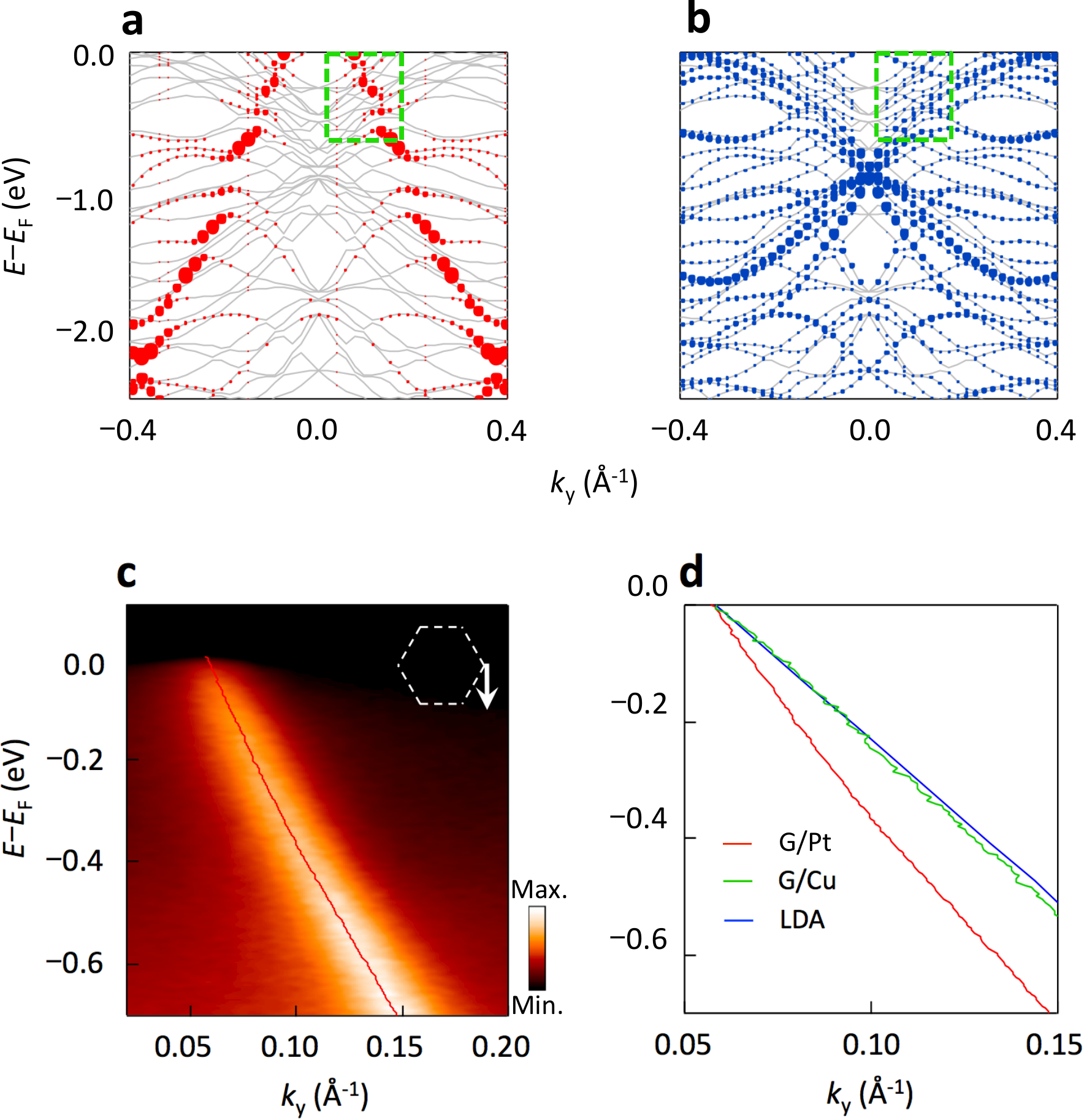}
\caption{ \label{Fig4}{\bf a-b}, Calculated electron band structure of graphene/Pt near $E_{\rm F}$ perpendicular to the $\Gamma{\rm K}$ direction of the graphene unit cell. Red and blue dots denote C 2$p_{\rm z}$ and Pt 5$d$ orbitals, respectively. {\bf c}, An ARPES intensity map near $E_{\rm F}$ taken perpendicular to the $\Gamma{\rm K}$ direction as denoted in the inset. The red curve is a Lorentzian fit to the intensity map. {\bf d}, The red curve is the $\pi$ band of graphene on the Pt substrate.  The green and blue curves are the measured $\pi$ band of graphene on a Cu substrate and the calculated $\pi$ band without substrate within LDA, respectively.}
\end{figure}

The electron band structure near $E_{\rm F}$ reveals another intriguing insight on the electronic properties of graphene on the Pt substrate. Figures~4{\bf a} and~4{\bf b} show calculated electron band structure of graphene/Pt taken perpendicular to the $\Gamma$K direction, in which red and blue dots denote C 2$p_{\rm z}$ and Pt $5d$ orbitals, and grey lines denote all the other orbitals. Near $E_{\rm F}$ denoted by the green dashed-rectangle, the graphene $\pi$ band shows weaker hybridization gaps compared to the one observed along the $\Gamma$K direction (Fig.~3) or at higher energies. In other words, the Pt 5$d$ bands shown in Fig.~4{\bf b} do not show a clear footprint of the hybridization with the $\pi$ band discussed in Figs.~3{\bf c} and~3{\bf d}. Indeed, an ARPES intensity map taken near $E_{\rm F}$ perpendicular to the $\Gamma$K direction (Fig.~4{\bf c}) exhibits not only continuous intensity distribution, but also an almost linear dispersion unlike the result taken along the $\Gamma$K direction (Figs.~3{\bf a} and~3{\bf b}) excluding a possibility of the hybridization with other states including impurity states.

The energy-momentum dispersion is extracted by a Lorentzian fit to each MDC as shown by the red curve in Figs.~4{\bf c} and 4{\bf d}. Interestingly, the observed slope of the dispersion is steeper than that of the calculated band within the local density approximation (LDA) (blue curve in Fig.~4{\bf d}). Typically when dielectric screening increases, the electron-electron interactions in graphene are strongly suppressed, so that the graphene band approaches towards the LDA band~\cite{HwangSR}. Indeed, the band structure of graphene/Cu is in good agreement with the LDA band as compared in Fig.~4{\bf d}. The difference between the measured band and the LDA band is a good approximation of the real part of electron self-energy~\cite{Siegel,HwangSR}. A logarithmic fit to the self-energy that is typically valid for charge neutral graphene~\cite{Siegel,HwangSR} gives an effective dielectric constant  $\epsilon=28.9$. Within the standard approximation of $\epsilon=(\epsilon_{\rm vacuum}+\epsilon_{\rm substrate})/2$, we obtain $\epsilon_{\rm substrate}=56.8$ for the Pt substrate. This is in excellent agreement with $58\pm10$ obtained by reflectivity measurements for a Pt film~\cite{Choi}. The finite effective dielectric constant of graphene indicates that the electron-electron interaction is not fully suppressed in graphene despite it stands on a metallic substrate, i.\,e.\,, a correlation effect in graphene/Pt is beyond the LDA can describe this system.

The observed hole doping and hybridization also differ graphene/Pt from all the other graphene on typical metals such as Cu~\cite{SiegelCu}, Ni~\cite{Varykhalov}, Co~\cite{Rader}, and Ru~\cite{Sutter}, which are electron-doped where $E_{\rm D}$ lies 0.3$\sim$2~eV below $E_{\rm F}$. Although weak hole doping of 0.10~eV and 0.06~eV has been observed from graphene on Ir(111)~\cite{Pletikosic} and Pt(111)~\cite{Shuyun} substrates, respectively, the graphene $\pi$ band is comparable to that of nearly freestanding graphene for these cases. On the other hand, another experimental studies on graphene on a Pt(111) substrate show that the graphene $\pi$ band exhibits a deeper hole-doping of 0.15~eV~\cite{Ilya1,Ilya2} and complex hybridization resulting in nontrivial spin structure of the $\pi$ band. Surprisingly, first-principles calculations on graphene/Pt(111)~\cite{Giovannetti,Wang} predict strong hybridization and hole doping as much as $\sim$0.5~eV, both of them excellently in agreement with our results.

It is interesting to note that one of the prominent differences of strongly interacting graphene with Pt from nearly freestanding graphene on Pt is the graphene growth method (see Supplementary Information for detailed discussion). Since the former using the CVD method shows hybridization and deeper hole-doping whether using polycrystalline Pt as shown in Figs.~2 and~3 or single crystal Pt(111)~\cite{Ilya1,Ilya2}, crystallinity of Pt does not play a crucial role in the observed strong interaction between graphene and Pt. On the other hand, the latter using carbon surface segregation~\cite{Shuyun,Sutter2} shows the absence of the hybridization and weaker hole-doping. Especially, for the case of higher temperature segregation, e.\,g.\,, 1600~$^{\circ}$C~\cite{Shuyun}, the electron band structure of the Pt substrate completely disappears. This possibly suggests that the surface morphology of the Pt substrate might become ill-defined by the carbon segregation. When the work function difference between graphene and Pt results in the hole doping~\cite{Giovannetti}, the ill-defined surface morphology of Pt will reduce its work function~\cite{Li}, leading to decreased hole doping concentration in graphene~\cite{Shuyun}. In addition, the hybridization should be also strongly influenced by the surface structure, consistent with the absence of the hybridization in the high quality epitaxial graphene prepared at higher temperature~\cite{Shuyun} compared to strong hybridization~\cite{Ilya1,Ilya2} as shown in Fig.~3.

Our results exemplify that the interface created between two-dimensional materials and transition metals provides a unique opportunity to engineer physical properties. By the formation of an interface with various substrates, not only the basic properties of graphene, such as charge carrier density~\cite{Pletikosic,HwangSR,Sprinkle,Siegel,Johansson,SiegelCu,Varykhalov,Rader,Sutter} and mobility~\cite{Dean}, can be tuned but also the complicated electronic correlation~\cite{Siegel,HwangSR}. Moreover, intriguing magnetic properties emerges such as a single spin Dirac cone~\cite{Usachov}, half-metallicity~\cite{Ilya2}, and spin-dependence variable range hopping~\cite{HwangS}, that are not observed when graphene stands alone. Such newly given functionalities of graphene open up the route towards the applications of graphene as a building block of a smart device with variable electron mobility within a single circuit controlled by the modification of a substrate and a spintronic device with controlled mobility.

\section{Conclusion}
In summary, we have reported the unique electronic properties of CVD-grown graphene on a Pt foil. The presence of the Pt substrate results in not only hole doping in graphene as much as 0.44~eV, but also hybridization gaps in the graphene $\pi$ bands consistent with the previous theoretical results~\cite{Giovannetti,Wang} and Raman spectroscopy studies~\cite{Qin}, but different from recent experimental reports on the epitaxial graphene on a Pt(111) substrate~\cite{Shuyun,Sutter2}. The combined study on the measured and calculated electron band structure singles out graphene/Pt as a unique system compared to all the other graphene on metallic substrates.

\acknowledgements
The authors thank E. H. Hwang for helpful discussions. This work was supported by the National Research Foundation of Korea (NRF) grant funded by the Korea government (MSIP) (No.~2015R1C1A1A01053065 and No.~2017K1A3A7A09016384). H.R. acknowledges support from the National Research Foundation of Korea (NRF) funded by the Ministry of Science, ICT and Future Planning (No.~2016K1A4A4A01922028). The Advanced Light Source is supported by the Office of Basic Energy Sciences of the U.S. Department of Energy under Contract No. DE-AC02-05CH11231. J. L. acknowledges support from the National Research Foundation of Korea (NRF) Grant funded by the Korean government (MSIP) (NRF-2015R1C1A1A01053810). A.L. acknowledges support from Berkeley Lab's program on sp2 bond materials, funded by the U.S. Department of Energy, Office of Science, Office of Basic Energy Sciences, Materials Sciences and Engineering Division, of the U.S. Department of Energy (DOE) under Contract No.~DE-AC02-05CH11231.

\end{document}